\documentclass[1 2pt]{article}
\usepackage{amssymb}
\begin{document}
\title{{A 2-Component Generalization of the Degasperis - Procesi Equation }}

\author{ Ziemowit Popowicz$^{a}$\\
\\
$^{a}$ Institute of Theoretical Physics\\
University of Wroc\l aw\\
pl. M. Borna 9, 50 -205 Wroc\l aw, Poland\\
e-mail: ziemek @ ift.uni.wroc.pl \\}

\maketitle

\hspace{4cm} {\bf Abstract: }
\vspace{1cm}

We present two different hamiltonian extensions of the Degasperis - Procesi equation to the two component equations. 
The construction based on the observation that the second Hamiltonian operator of the Degasperis - Procesi 
equation could be considered as the Dirac reduced Poisson tensor of the second Hamiltonian operator of the 
Boussinesq equation. The first extension is generated 
by the Hamiltonian operator which is  a Dirac reduced operator of the generalized but degenerated 
second Hamiltonian operator of the Boussinesq equation. The second one is obtained by the  $N=2$ supersymmetric 
extension of the mentioned method. As the byproduct of this procedure  we 
obtained the Hamiltonian system of interacting  equations which contains the Camassa - Holm and Degasperis - 
Procesi equation.

\newpage

\section*{Introduction}

Recently a family of equations of the form [1 - 7] 
\begin{equation}
u_t - u_{xxt}  = \frac{1}{2} \big ( -(b+1) u^2 + 2uu_{xx} + (b-1) u_x^2 \big )_x
\end{equation}
has been investigated in the literature . 

When $ b=2$ Eq.(1) reduces to the Camassa - Holm 
\begin{equation}
u_t - u_{xxt}  = \frac{1}{2} \big ( -3 u^2 + 2uu_{xx} +  u_x^2 \big )_x
\end{equation}
equation, which  describes a special approximation of shallow water theory. This equation shares most of the 
important properties of an integrable system of KdV type, for example, the existence of Lax pair formalism, the 
Bi-Hamiltonian structure, the multi-solitons solutions. Moreover this equation admits peaked solitary wave solutions.

Degasperis and Procesi showed that the Eq. (1)  is integrable also for the $b=3$ case. 
The Degasperis - Procesi equation 
\begin{equation}
u_t - u_{xxt} = (-2u^2 + uu_{xx} + u_x^2 )_x 
\end{equation}
can be considered as a model for shallow - water dynamics also  and found to be completely integrable. 
Similarly to the Camassa - Holm case the Degasperis - Procesi equation has the Lax pair and admits 
peakon dynamics also. 

The Camassa - Holm and Degasperis - Procesi possesses the Bi - Hamiltonian structure and recursion operators. 
In the case of the Camassa - Holm equation there are two local Hamiltonian structures, given by 
\begin{eqnarray}
&& \hspace{3cm}  m_t =  B_o \frac{\delta H_2}{\delta m} = B_1\frac{\delta H_1}{\delta m}  \\ \nonumber 
&& B_o =  - \partial ( 1 - \partial^2) = - {\cal L} , \hspace{2cm} B_1 = -( m\partial + \partial m) \\ \nonumber 
&& H_2 = \frac{1}{2} \int  dx (u^3 + uu_x^2), \hspace{2cm} H_1 = \frac{1}{2} \int  dx (u^2 + u_x^2)
\end{eqnarray} 
with $m=u-u_{xx}$, whose compatibility was known in {\cite{fuch}}. 
In the case of the Degasperis - Procesi equation, there is only one local Hamiltonian structure and the 
second Hamiltonian structure is nonlocal 
\begin{eqnarray}
&& \hspace{3cm} m_t =  B_o \frac{\delta H_{-1}}{\delta m} = B_1\frac{\delta H_o}{\delta m}  \\ \nonumber 
&& B_o = {\cal L}(4-\partial^2), \hspace{2cm} B_1=(m_x +3m\partial) {\cal L}^{-1} 
(2m_x + 3m\partial) \\ \nonumber 
&& H_{-1} = -\frac{1}{6}\int  u^3 dx ,\hspace{2cm}  H_o=-\frac{1}{2} \int  m dx
\end{eqnarray}
whose compatibility was proven in {\cite{hone1}}.

The two component generalization of Camassa - Holm equation 
\begin{eqnarray}
m_t & = & -um_x - 2mu_x + \rho \rho_x \\ \nonumber 
\rho_t & = & - (\rho u)_x, {\label{general}}
\end{eqnarray}
where $m=u-u_{xx}$, has been recently proposed by Chen, Liu and Zhang  {\cite{chiny} and Falqui {\cite{falqui}}. 
This generalization, similarly to the Camassa - Holm equation, is 
the first negative flow of the AKNS hierarchy and possesses the peakon and multi - kink 
solutions  and possesses the Bi-Hamiltonian structure {\cite{chiny, heniek, falqui}}. The basic idea of this 
generalization was to include the additional function to the Lax pair and then to extract the basic properties 
of the equation from this generalized Lax pair representation.

In this paper we show that it is possible to construct two different generalizations of the two 
component version of the Degasperis - Procesi equations as the Hamiltonian equations  in the form:
\begin{eqnarray}
\rho_t &=& -k_2\rho_x u -(k_1+k_2) \rho u_x \\ \nonumber 
m_t &=& - 3mu_x -  m_xu + k_3\rho \rho_x 
\end{eqnarray} 
where $m=\partial^{-1}{\cal L}u$ while $k_1=k_2=1$ and $k_3$ is an arbitrary  constant or 
$k_2=1,k_3=0$ and $k_1$ takes an  arbitrary value

The second generalization  is 
\begin{eqnarray}
\rho_t &=& -2\rho  u_{x}  - \rho_x u   \\ \nonumber 
m_t &=& - 3mu_{x} -   m_xu - \rho u_{x}  + 2\rho\rho_x
\end{eqnarray}
where $m=\partial^{-1}{\cal L}u$

We show also  that it is possible to construct the interacting system of equations which 
contains the Camassa - Holm and  Degasperis - Procesi equations 
\begin{eqnarray}
m_{t}  &=&  - 3m(2u_x + v_x) - m_x(2u+v)\\ \nonumber 
n_{t}  &=& - 2n(2u_x + v_x) - n_x(2u+v).
\end{eqnarray} 
where $m=u-u_{xx}, n=v-v_{xx}$.

The construction presented in this paper based on the generalization of the second Hamiltonian 
operator $B_1$ of the Degasperis - Procesi to the the two dimensional  matrix operator.
The direct manner of the generalization leads us to very complicated assumptions on the entries 
of the matrix and very painful verifications of the Jacobi identity. We omit these complications in three 
steps. In the first step, which we will call as the decompression of the Hamiltonian operators,  
we consider the  Hamiltonian pencil $B_o + B_1$ of the Degasperis - Procesi equation, 
as the Dirac reduced operator of the second Hamiltonian operator of the Boussinesq equation. 
As the result  we obtained the two dimensional local matrix Hamiltonian operator. 
In the  second step we generalize this  operator to the three dimensional  matrix operator in such 
the way that to  fulfill  the Jacobi identity. In the last step we apply the Dirac reduction with respect to 
the decompression function appearing in the first step.

The paper is organized as follows. In the first section we describe the Dirac reduction technique and we
show that the Hamiltonian pencil of the Camassa - Holm equation follows from the second Hamiltonian 
operator of the nonlinear Schr{\"o}dinger equation. In  the second section we decompress the Hamiltonian 
operator of the Degasperis - Procesi equation to the second Hamiltonian operator of the Boussinesq equation.
In the third section we carry out the Dirac reduction of the generalized but degenerated Hamiltonian operator 
of the Boussinesq equation. The first two component generalization of the Degasperis - Procesi equation is 
presented in the third section. The fourth section contains the description of the interacting system of 
Camassa - Holm and Degasperis - Procesi equations. The fifth section contains the supersymmetric investigation
of decompression method which allowed us to obtain second generalization of the two - component 
Degasperis - Procesi equation. The last section contain concluding remarks.

\section{The Dirac Reduction of the Poisson tensor.} 
The energy - dependent Schr{\"o}dinger spectral problem {\cite{ford}} for Camassa - Holm equation can be formulated with the help 
of Lax operator as
\begin{eqnarray}
\Psi_{xx} &=& (\frac{1}{4}  - \lambda m) \Psi \\ \nonumber 
\Psi_t &=& - (\frac{1}{2\lambda} + u )\Psi_{x} + \frac{1}{2}u_x\Psi
\end{eqnarray}
 Compatibility condition for the above system, yields two independent equations 
\begin{eqnarray}
m_t &=& -2mu_x -m_xu \\ \nonumber 
m &=& u - u_{xx}
\end{eqnarray}
We would like to obtain the Hamiltonian operator for the Camassa - Holm    equation   and therefore we 
consider a more general  system then (10)
\begin{eqnarray}
\Psi_{xx} &=& v\Psi_x - w \Psi \\ \nonumber 
\Psi_t &=& - A_1\Psi_{x} + A_2\Psi
\end{eqnarray}
where now $w,v$ are given functions while $A_1,A_2$ are at the moment arbitrary functions.
This system can be reduced to the Lax representation (10) for the special choice of the functions  
$w,v,A_1,A_2$ and if we additionally assume the dependence on the spectral parameter. 
The compatibility conditions for  Eq. (12) gives us the following time evolution of the 
functions $v,w$
\begin{equation}
\left(
\begin{array}{c}
v \\ 
\noalign{\vskip 4pt}%
 w
\end{array}
\right)_t = J \left (
\begin{array}{c} A_2 \\ 
\noalign{\vskip 4pt}%
-A_1
\end{array} 
\right) =
\left(
\begin{array}{cc}
2\partial  &  \partial v + \partial^2  \\  
\noalign{\vskip 4pt}%
-\partial^2 +v\partial   & \partial w + w \partial 
\end{array} \right)
\left (
\begin{array}{c} A_2 \\ 
\noalign{\vskip 4pt}%
-A_1
\end{array} 
\right),
\end{equation}
To establish the Hamiltonian character of the corresponding flows (13) we have to choose $A_1$ and $ A_2$ 
in such a way that they  constitute the coordinates of variational derivatives of some functionals ${\cal H}$.
However we follow in different way. Let us notice that our $J$ operator is the second Hamiltonian operator 
connected with the AKNS equations. Indeed under the "coordinate change"  
\begin{equation}
v=\frac{q_x}{q} \hspace{2cm} w=pq
\end{equation}
the $J$ operator transforms to 
\begin{equation}
 J  =
\left(
\begin{array}{cc}
-2 p\partial^{-1} p &  \partial + 2p\partial^{-1}q  \\  
\noalign{\vskip 4pt}%
\partial +2q\partial^{-1}p    & -2q\partial^{-1} q  
\end{array} \right)
\end{equation}
A perhaps less known fact is the following. Under the Dirac reduction where $q=1$ or $p=1$ this 
Hamiltonian reduces to the second Hamiltonian operator for the Korteweg - de Vries equation.  
We use the standard reduction lemma for Poisson brackets {\cite{walter}} which can be stressed as:
For the given Poisson tensor 
\begin{equation}
P(v,w)= \left(
\begin{array}{cc}
P_{vv}(v,w) &  P_{vw}(v,w) \\  
\noalign{\vskip 4pt}%
P_{wv}(v,w) & P_{ww}(v,w)
\end{array} \right)
\end{equation}
let us  assume  that $P_{vv}(v,w)$ is invertible, then for arbitrary $v$  the map given by 
\begin{equation}
\Theta(w;v)= P_{ww}(v,w) - P_{wv}(v,w) \big( P_{vv}(v,w) \big)^{-1}P_{vw}(v,w) 
\end{equation}
is a Poisson tensor where $v$ enters the reduced Poisson tensor $\Theta$ as a parameter rather than 
as a variable. The reduced Poisson tensor $\Theta(v:w)$ reads 
\begin{equation}
\Theta(v;w)= P_{vv}(v,w) - P_{vw}(v,w) \big( P_{ww}(v,v) \big)^{-1}P_{wv}(v,w) 
\end{equation}

Now we can apply this reduction to the Hamiltonian operator defined in  Eq.(13) where we assume that $v=1$ and
as the result we obtain the following operator
\begin{equation}
\Theta = -\frac{1}{2} (\partial - \partial^3) +\partial w + w\partial 
\end{equation}
It appear that this operator is the linear combinations of our first and  second Hamiltonian operators 
of the Camassa - Holm equation .

On the other side we can carry out the Dirac reduction with respect to the function $w$ where now $w=1$. 
As a result we obtained the  following Poisson tensor
\begin{equation}
\Theta = 2\partial - \frac{1}{2}\partial^3 - \frac{1}{2} \partial v \partial^{-1} v \partial 
\end{equation}
which is the linear combination of the first and second Hamiltonian operators of the 
Modified Korteweg - de Vries equation.

\section{The Decompression of the Hamiltonian pencil of the Degasperis - Procesi equation.} 

The energy-dependent Lax operator responsible for the Degasperis - Procesi equation is {\cite{dega1,dega2}}
\begin{eqnarray}
\Psi_{xxx} &=& \Psi_x - \lambda m\Psi \\ \nonumber 
\Psi_t &=& - \lambda^{-1}\Psi_{xx} -  u \Psi_x +u_x\Psi 
\end{eqnarray}
where $\lambda$ is the parameter. The compatibility conditions  gives us the 
Degasperis - Procesi equation.

In order to obtain  the Hamiltonian operator for the Degasperis - Procesi equation we 
consider more general system of equations then (21) 
\begin{eqnarray}
\Psi_{xxx} &=& v \Psi_x +\frac{1}{2} (v_x + 2z) \Psi \\ \nonumber 
\Psi_t &=& A\Psi_{xx} + (A_1 - \frac{1}{2} A_x)  \Psi_x +\frac{1}{6}(A_{xx} - 6A_{1,x} -4vA)\Psi 
\end{eqnarray}
where $v,z$ are given functions while $A,A_1$ are at the moment arbitrary functions. 
This system can be reduced to the Lax representation (21) for the special choice of the functions  
$v,z,A,A_1$ and if we additionally assume the dependence on the spectral parameter. 

The compatibility conditions for  Eq. (22) gives us 
\begin{equation}
\left(
\begin{array}{c}
v \\ 
\noalign{\vskip 4pt}%
 z
\end{array}
\right)_t = J \left (
\begin{array}{c} A_1 \\ 
\noalign{\vskip 4pt}%
A
\end{array} 
\right) =
\left(
\begin{array}{cc}
-2\partial^3 + 2v\partial + v_x &  3z\partial + 2z_x  \\  
\noalign{\vskip 4pt}%
3z\partial +z_x  & \frac{1}{12} J_{2,2}
\end{array} \right)
\left (
\begin{array}{c} A_1 \\ 
\noalign{\vskip 4pt}%
A
\end{array} 
\right),
\end{equation}
where 
\begin{equation}
J_{2,2}= 2\partial^5 - 10v\partial^3-15v_x\partial^2 +(8v^2-9v_{xx})\partial + 8v_xv - 2v_{xxx} 
\end{equation}
We recognize that $J$ is the second Hamiltonian operator for the Boussinesq equation. 
We can easily obtain the Boussinesq equation assuming $ A_1 = 0, A=1$  what gives us 
\begin{eqnarray}
v_t & = & 2z_x \\ \nonumber 
z_t & = &  \big ( -2v_{xxx} + 8v_xv \big )/ 12
\end{eqnarray}  
Let us now investigate the behavior of the $J$  operator under the Dirac reduction where now we assume  $v=1$.  
Using the formula (17) we obtained the following Poisson tensor 
\begin{equation}
B = \frac{1}{6}\partial(4 - \partial^2)(1 - \partial^2 ) -\frac{1}{2}(3z\partial + z_x)
(\partial -\partial^3)^{-1}(3z\partial + 2z_x) 
\end{equation}
We quickly  recognize,  after the identification $z=m$,  that $B$ is the linear combination of the first and 
second Hamiltonian operators of the Bi-Hamiltonian structure of the Degasperis - Procesi equation. 
Thus the $B$ operator 
\begin{equation} 
B=\frac{1}{6}B_o -\frac{1}{2}B_1
\end{equation}
satisfy the Jacobi identity due to the Dirac reduction. Moreover using the scaling argument to the function $z$ 
we can easily verify  that $B_o$ and $B_1$  are the compatible operators and that $B_1$ satisfy the Jacobi 
identity as well. 

This  process we will call as the decompression of the Hamiltonian structure. More precisely  in this 
procedure we try to find higher dimensional operator  for which the Dirac reduction gives us the 
Hamiltonian operator under consideration. 
In some sense it is an inverse operation to the Dirac reduction technology. 
The advantage of this decompression technique is a possibility of quick verification of  the Jacobi identity 
for the nonlocal Hamiltonian operators. Indeed if we embedd 
some nonlocal  Hamiltonian operators, in such  a way,  that the final operator will be local, 
then it is much easier to check the Jacobi identity  compare to the  verification of this identity for the 
nonlocal  operators.  For example, the decompressed second Hamiltonian operator of the AKNS equations Eq.15, 
gives us the local Hamiltonian operator which is connected with the Kac - Moody $sl(2)$ algebra {\cite{pops}}.
The disadvantage of this technique is a lack of uniqueness because  we can embed 
the given Hamiltonian operator to the higher dimensional matrix operator in the different manner also as 
we see in the next section.
 
It is hard   to define the general prescription of the decompression procedure using the examples 
mentioned earlier. In the case of the nonlocal Hamiltonian operators this construction requires  many 
assumptions if we would like to obtain, as the final result, the higher dimensional local operator. However 
in this paper  we  would like extend the $J$ operator defined in Eq. 23 to the three dimensional Hamiltonian 
matrix operator including new function, in such a way that to preserve the gradation of the matrix 
elements with respect to the weights of the functions. In the next step we carry out the Dirac reduction for 
this extended matrix operator when $v=1$ and obtain some  Hamiltonian opreator. Then we use  this  
new Hamiltonian operator to the construction of the two - component  Degasperis - Procesi equations. 

To finish this section let us notice that it is possible to consider  more general form of the 
Hamiltonian  operator then this  defined by the Eq. 23. Indeed let us  consider the following operator  
\begin{equation}
{\cal J} =\left(
\begin{array}{cc}
c\partial^3 + 2v\partial + v_x &  3z\partial + 2z_x  \\  
\noalign{\vskip 4pt}%
3z\partial +z_x  & {\cal J}_{2,2}
\end{array} \right )
\end{equation}
where $c$ is an arbitrary central extension term and ${\cal J}_{2,2}$ is constructed out of 
the function $v$ its derivatives and from the differential operators only. The  verification of 
the Jacobi identity leads us to the conclusion that this identity holds if 
${\cal J}_{2,2}$ is defined as
\begin{equation}
{\cal J}_{2,2} =  \frac{\lambda}{16}(c^2\partial^5 + 10cv\partial^3 + 15cv_x\partial^2 + 
(9cv_{xx} + 16v^2 ) \partial + 2cv_{xxx} + 16v_xv)
\end{equation}
where $\lambda$ is an arbitrary constant.  We see that $\cal J$ reduces to the $J$ operator 
when $c=-2$ and $\lambda=\frac{2}{3}$.
The most interesting case is the degenerated one where we assume that $ \lambda=0$ 
while $c$ is an arbitrary constant. For this  degenerated case, it is possible also,  to construct the 
second Hamiltonian operator for the  Degasperis - Procesi equation using the decompression described earlier. 
However then the information on the first Hamiltonian operator is lost.

\section{A Two Component Degasperis - Procesi equation.}

\hspace{0.5cm}  Let us decompress  the $\cal J$ operator defined by Eq. (28) to the three  dimensional matrix $\Gamma$ 
including  the new function $\rho$ which has the same weight as the $v$ function. 
We try to find the  general form of  this  operator 
in such a way,  that  to preserve the gradation  of the matrix elements with respect to the weights of the functions.

Let us first consider the decompression for  the degenerated case where $\lambda=0, c=-2$ and   $z=m$. 
We make the following assumptions on the entries of the $\Gamma$ matrix 
 $\Gamma_{1,1}=J_{1,1}, \Gamma_{1,2}=J_{1,2},\Gamma_{2,1}= J_{2,1}$ 
where  $J_{1,1},J_{1,2},J_{2,1}$ are defined in eq. (23). For the rest elements we assumed that they 
are  constructed  out of  the function $\rho$ its derivatives and differential operators in 
such a way that  they reduces to $0$ when $\rho=0$.
As we checked, using the computer algebra, that the following matrix 
\begin{equation}
\Gamma =
\left(
\begin{array}{ccc}
-2\partial^3 + 2v\partial + v_x & 3m\partial + 2m_x  & k_1\partial \rho + k_2 \rho\partial  \\ 
\noalign{\vskip 4pt}%
3m\partial +m_x  & \frac{1}{2}k_3 \rho \partial \rho & 0 \\ 
\noalign{\vskip 4pt}%
k_2\partial \rho + k_1 \rho \partial & 0 & 0  \end{array} \right)
\end{equation}
satisfy the Jacobi identity  for the two choices of free parameters:
 $ k_1=k_2=1$ and $k_3$ is an arbitrary value for the first case while 
$ k_2=1,k_3=0$ and $k_1$ is an arbitrary value for the second case. 

If we, similarly to the previous case,  carry out 
the Dirac reduction with respect to the $v=1$  we obtain the following matrix Hamiltonian operator 

\begin{equation}
{\cal Z}  =
-\frac{1}{2} \left(
\begin{array}{cc}
\begin{array}{c}
  (3m\partial +m_x){\cal L}^{-1}(3m\partial +2m_x) \\
- k_3\rho\partial \rho \end{array} &
  (3m\partial +m_x){\cal L}^{-1} (k_1\partial \rho  +k_2\rho \partial) \\
\noalign{\vskip 4pt}%
  (k_2\partial \rho +k_1 \rho \partial) {\cal L}^{-1}(3m\partial +2m_x) &
  (k_2\partial \rho  +k_1 \rho ) {\cal L}^{-1} (k_1\partial \rho  +k_2\rho \partial) \\ 
 \end{array} \right)
\end{equation}

Let us compute the equation of motion for the Hamiltonian $H= \int dx m $. Assuming that $ m=u - u_{xx} $ we 
obtain 
\begin{eqnarray}
m_{t}  &=&  k_3\rho \rho_x/2 - 3mu_x -m_xu \\ \nonumber 
\rho_t &=& -k_2\rho_x u -( k_1 +k_2) \rho u_x.
\end{eqnarray} 
It is our two - component generalization of the Degasperis - Procesi equation.
In the case $k_3=0$, the two equations (32) are no more coupled and the equation on $\rho$ becomes linear.

We tried  to find the first Hamiltonian operator for the system (34), decompressing   
the $\cal J$ operator defined in eq. (28)  in   the nondegenrated case .
Unfortunately we have been not able to find any such  operator  and moreover we 
did not defined any new generalization of the Degasperis - Procesi equation in that manner. 
We make the same assumptions on the decompressed matrix $\hat \Gamma$ as previously with the one restriction.
The  element $\hat \Gamma_{2,2}$  is  constructed out of the functions $v,\rho$ its 
derivatives and differential operators and  it reduces to ${\cal J}_{2,2}$ when $\rho=0$. 
We fixed the $\hat \Gamma$ matrix  verifying  the Jacobi identity, carry out the Dirac reduction of 
the $\hat \Gamma$ matrix  with respect to the function $v=1$ and we  obtained the reduced matrix in the form.
${ \Omega}  = {\cal Z}_o + {\cal Z}$ ,
where ${\cal Z}$ is defined by eg. (31). 
 The computer algebra identified only one  operator  ${\cal Z}_o$ which however as we 
checked  does not satisfy the Jacobi identity. It means that ${\cal Z}_o$ is not the first Hamiltonian 
operator for the system (32). 

We have been not able to find any additional constants of motion for the system (32). We tried to find these 
constants from the Lax representation. Therefore we verified  two different assumptions  on the  Lax operator 
which should gives us the two - component  Degasperis - Procesi equations. 
The first was the matrix generalization of the Lax operator responsible for the Degasperis - Procesi equation 
while in  the second we assumed the polynomial dependence of the spectral parameter.

 Unfortunatelly we did not  find any  Lax representation for the two - component Degasperis - Procesi equations 
and hence we did not established the integrability of the system in that manner. 
However it does not mean that this system is not integrable. 
We need quite different methods in  order to establish the integrability of the system (34) as for example to 
try to find the recursion operator. It seems that the  problem of the existence of 
higher order constants of motion and the recursion operator for the system (34) is worth to study.

\section{Degasperis - Procesi Equation interacted with Camassa - Holm Equation}

Let us consider the case when $k_3=0$ and $k_1=k_2=1$ and 
redefine the variables as $\rho = n=v - v_{xx}$, then the Hamiltonian operator $2{\cal Z}$ 
defines  new Hamiltonian operator 
\begin{equation}
 Z  =
-\left(
\begin{array}{cc}
  9 m^{2/3} \partial m^{1/3} {\cal L}^{-1} m^{1/3}\partial m^{2/3} &
  6 m^{2/3}\partial m^{1/3} {\cal L}^{-1} n^{1/2}\partial n^{1/2} \\
\noalign{\vskip 4pt}%
  6  n^{1/2}\partial n^{1/2} {\cal L}^{-1}  m^{1/3}\partial m^{2/3} &
  4 n^{1/2}\partial n^{1/2} {\cal L}^{-1}  n^{1/2}\partial n^{1/2} \\ 
 \end{array} \right)
\end{equation}
where ${\cal L}^{-1} = \partial^{-1}(1-\partial^2)^{-1}$. This operator  satisfy the Jacobi identity due to 
the Dirac reduction of the $\Gamma$ operator defined by Eq. 30. Thus one can define 
the following equations of motion 
\begin{eqnarray}
m_{t}  &=&  - 3m(2u_x + v_x)   - m_x(2u+v)  \\ \nonumber 
n_{t}  &=&  - 2n(2u_x + v_x) - n_x(2u+v).
\end{eqnarray} 
if we apply the $Z$  operator to the Hamiltonian  $H=\int dx (m+n) $ . 
It is our interacting system of equations which contains the Camassa - Holm and Degasperis - Procesi 
equations. 
Indeed  when $n=v=0$ and we rescale the time our system reduces to the Degasperis - Procesi equation while the 
reduction $ m=u=0$ leads us to the Camassa - Holm equation. 

For our best knowledge it is a new system of equations and one can ask whether this 
system is integrable.  We have found three independent conserved quantities
\begin{eqnarray}
&& H_0 = \int (m+n) dx \\ \nonumber 
&& H_1 = \int n^{\lambda} m^{(1-2\lambda)/3} dx \\ \nonumber 
H_2 &=& \int (-9n_x^2 n^{\lambda -2} m^{-(1+2\lambda)/3} + 12 n_xm_x n^{\lambda -1}m^{-(4+2\lambda)/3} - 
4m_x^2n^{\lambda} m^{-(7+2\lambda)/3} ) dx
\end{eqnarray}
where $\lambda$ is an arbitrary constant. The $H_1$ conserved quantity is the Casimir function for our 
Hamiltonian operator Eq. (33). Interestingly when $\lambda =0$ then $H_1$  reduces to the Casimir function 
for the Degasperis - Procesi equation while for $\lambda = 1/2 $ it reduces to the Casimir function for the 
Camassa - Holm equation. The existence of three independent conserved quantities is a good sight to 
expect that this system is integrable. 

The popular manner of checking the integrability is to define the recursion operator using the Bi-Hamiltonian formulation.
However we could not find such structure. On the other side the most easy  manner of verifying  the integrability is to 
define the Lax representation for the given partial differential equation. If such representation exists for our system  (34) 
then this  should be reduced to the Degasperis-Procesi  or to the Camasaa-Holm Lax representation 
when $n=v=0$ or $m=u=0$ respectively. One can to think therefore, that the system of interacting Camassa - Holm and  
Degasperis - Procesi equations appear as  the multi-component generalization of the Lax operator responsible for 
the Degasperis - Procesi equation. It is well known that an extension of a scalar integrable partial differential 
equations to a multi-component version,  as for example for the  vector nonlinear Schr{\"o}dinger equation,   
are still   integrable and  can be achieved  by considering the corresponding Lax pair in a higher rank matrix algebra.
We verified such possibility and  therefore we considered the most general assumption on the  two dimensional 
matrix generalization of the  Lax operator of the Degasperis - Procesi equation which contained  also the Lax operator 
of the Camasaa - Holm equation. However we did not find any operator  which produces the system (34). 
The difficulties in such construction are connected probably with the different orders of the differential 
operators in the  Lax operators of the Camassa - Holm and Degasperis - Procesi equations.

The next  possibility of checking the integrability, is to consider the  third order energy-dependent scalar Lax 
operator where the polynomial dependence of the spectral parameter is assumed {\cite{afl}}. It was shown 
in {\cite{chiny,heniek}} that if one 
allows the polynomial dependence of the scalar parameter for the second order energy-dependent scalar Lax 
operator then this leads us to the two component generalization of the Camassa-Holm equation. We have checked, that
the same strategy can not be applied for the Degarsperis-Procesi equation.

\section{The Extended N=2 Supersymmetric Degasperis - Procesi Equation.}

We will use now the supersymmetric formalism {\cite{wess}} which allows us to consider  the supersymmetric 
analog of the second Hamiltonian operator  which is connected with the degenerated second Hamiltonian 
operator of the Boussinesq equation. 
Here we will use the supersymmetric algebra of (super) derivatives where 
\begin{eqnarray}
{\cal D}_1 &=& \frac{\partial}{\partial \theta_1 } - \frac{1}{2}\theta_2\frac{\partial}{\partial x} \hspace{2cm} 
{\cal D}_2 = \frac{\partial}{\partial \theta_2} - \frac{1}{2}\theta_1\frac{\partial}{\partial x} \\ \nonumber 
 \{ {\cal D}_1,{\cal D}_2  \} & = & -\partial,  \hspace{0.5cm} [ {\cal D}_1,{\cal D}_2 ] =  
{\cal D}_1{\cal D}_2  -  {\cal D}_2,{\cal D}_1 , \hspace{0.5cm}  
  {\cal D}^2_1 = {\cal D}^2_2 =0 
\end{eqnarray}
The super functions can be thought as the $N=2$ supermultiplets which depends on $x$ and additionally on two 
Grassman valued functions with the entries
\begin{equation}
u(x,\theta_1,\theta_2) = u_o(x) + \theta_1\chi_1(x)  + \theta_2\chi_2(x) +\theta_2\theta_1 u_1(x)
\end{equation}
where $u_o(x), u_1(x)$ are the classical functions, $\chi_1(x),\chi_2(x)$ are Grassman valued functions 
and $\theta_1,\theta_2$ are the Mayorana spinors {\cite{wess}}. 

The main idea of the supersymmetry is to treat boson and fermion operators equally. In order to get 
supersymmetric theory we have to add to a system of $k$ bosonic equations $kN$ fermions and $k(N-1)$ boson fields 
$k=1,2,..N, N=1,2,..$ in such a way that the final theory becomes supersymmetric invariant. From the 
soliton point of view we can distinguish two important classes of the supersymmetric equations: the non-extended 
$(N=1)$ and extended $( N>1 ) $ cases. Consideration of the extended case may imply new bosonic equations whose 
properties need further investigation. 

There are many different methods of the supersymmetrization of 
the classical equations and many new integrable equations {\cite{pop1,pop2,mathie,delduc}} has been discovered 
in that manner. For example recently Devchand and Schiff {\cite{devchand}} found nonextended supersymmetric 
generalization of the Camassa - Holm eqution. The present author showed {\cite{pop0}} that extended $N=2$ 
supersymmetric generalization of the Camassa - Holm leads us directly to the two component 
generalization of this equation considered in {\cite{falqui,chiny}}.

The $N=2$ supersymmetric Boussinesq equation has been constructed utilizing the $N=2$ supersymmetric 
extension of the $W_3$ algebra {\cite{krzywy,krzywy1}}. 
This supersymmetric algebra  is generated by two $N=2$  supermultiplets, with the 
conformal spins $(1,\frac{3}{2},\frac{3}{2},2)$ and $(2,\frac{5}{2},\frac{5}{2},3)$ and exists at an arbitrary 
value of the central charge and  is connected with the following supersymmetric 
matrix operator $\hat J$  with the entries 
\begin{eqnarray}
\hat J_{1,1} &=& c[{\cal D}_1,{\cal D}_2]\partial + u_x  + 
u\partial +({\cal D}_1 u){\cal D}_2 + ({\cal D}_2 u){\cal D}_1 \\ \nonumber 
\hat J_{1,2} &=& 2\partial v + ({\cal D}_1 v){\cal D}_2  +({\cal D}_2 v){\cal D}_1 \\ \nonumber 
\hat J_{2,1} &=& \partial v + v\partial +  ({\cal D}_1 v){\cal D}_2 + ({\cal D}_2 v){\cal D}_1 .
\end{eqnarray}
where $c$ is an arbitrary  constant and  the element $\hat J_{2,2}$ has rather 
complicated form {\cite{krzywy,pop3} }. For the next purposes we assume 
that $c=-1$.

However we do not use this operator for our considerations because if we carry out the Dirac reduction 
with respect to $u=1$ it appears that $\hat J_{2,2}$ does not satisfy the Jacobi 
identity. On the other side the supersymmetric extension of $W_3$ algebra is unique, when $\hat J_{2,2} \neq 0$, 
so we restrict the consideration to the degenerated case where 
\begin{equation}
\hat J_{2,2} = 0
\end{equation}
The $\hat J$ matrix operator given by previous equations define proper Hamiltonian operator  what can be easily  
checking computing the Jacobi identity. 
We can apply the Dirac reduction scheme in the supersymmetric case as well. Let us carry out this 
reduction  where $u=1$ obtaining 
\begin{equation} 
\Theta =- \Big ( \partial v + v\partial + ({\cal D}_1 v){\cal D}_2 + ({\cal D}_2 v){\cal D}_1\Big) \  
\hat {\cal  L} \  
\Big ( 2\partial v + ({\cal D}_1 v){\cal D}_2 + ({\cal D}_2 v){\cal D}_1 \Big ) 
\end{equation}   
where 
\begin{equation}
\hat {\cal L} = \big (\partial  - [ {\cal D}_1,{\cal D}_2 ] )^{-1} = \partial^{-1}(1-\partial^2)^{-1} 
(1 +  [ {\cal D}_1, {\cal D}_2 ]) = {\cal L}^{-1} (1 +  [ {\cal D}_1, {\cal D}_2 ]).
\end{equation} 
This reduced operator generates the  following equation of motion when it acts on the  Hamiltonian 
$H =\frac{1}{2}\int dx d\theta_1 d\theta_2 v$  
\begin{equation}
v_t = -(2vA_x + v_xA +({\cal D}_1v)({\cal D}_2A)+ ({\cal D}_2v)({\cal D}_1A))
\end{equation}
where $A=\partial \hat {\cal L} v$ .

It is our supersymmetric extension of the Degasperis - Procesi equation. 
We have been not able to find any supersymmetric Lax representation responsible for this 
supersymmetric equation. 

Let us compute the bosonic sector, where  all fermionic components disappear, it means that  we 
consider  the superfunctions  in the form 
\begin{eqnarray}
A &=&  A_o + \theta_2\theta_1 A_1  \\ \nonumber 
v &=& V_o + \theta_2\theta_1 V_1 = (A_o -2A_1) + \theta_2\theta_1 (A_1 - A_{o,xx}/2).
\end{eqnarray}

In these coordinates we have 
\begin{eqnarray}
V_{o,t} &=&  - 2V_oA_{o,x}  - V_{o,x}A_o  \\ \nonumber 
V_{1,t} &=& - 3V_1A_{o,x} - V_{1,x}A_o - 2V_oA_{1,x}.
\end{eqnarray}

In order to have the connection with the Degasperis - Procesi equation let us introduce new variables $\rho$ and
$u$ 
\begin{equation}
V_o = \rho , \hspace{0.7cm} V_1 = \frac{1}{2} ( m - \rho ) , \hspace{0.7cm} m= u - u_{xx}, 
\end{equation}
in which 
\begin{equation}
A_0=u \hspace{1cm} A_1=\frac{1}{2}(u-\rho).
\end{equation}
Then  the equation (44) transforms to 
\begin{eqnarray}
\rho_t &=& - 2\rho  u_{x}  - \rho_x u   \\ \nonumber 
m_t &=& - 3mu_{x} -  m_xu - \rho u_{x} + 2\rho\rho_x
\end{eqnarray}
In that manner we obtained the second two component generalization of the Degasperis - Procesi equation.

\section*{Conclusion.}
 In this paper we considered two different extensions of the Degasperis - Procesi equation. Our  construction based 
on the observation that the second Hamiltonian operator of the Degasperis - Procesi equation could 
be considered as the Dirac reduced Poisson tensor of the second Hamiltonian operator of the Boussinesq equation.
The first extension is generated by the Hamiltonian operator which is obtained as a Dirac reduced operator of the 
generalized but degenerated second Hamiltonian operator of the Boussinesq equation. The second one is generated 
by the supersymmetric $N=2$ extension. Unfortunatelly we did not  find any  Lax representation for the two - component Degasperis - Procesi equations 
and hence we have been not able to verify the integrability of the systems. 
We also presented the interacting system of equations 
which contains the Camassa - Holm and Degasperis - Procesi equation. For this interacting system we constructed 
few conserved quantities. 
 The decompression method presented here does not allow to construct the 
first Hamiltonian operators for our systems,  because this method  based on the extensions of some local 
Hamiltonian operators which eliminate this structure from the very beginning. However it does not mean that 
this structure does not exist.  If  the first Hamiltonian structures  appear  in  our systems, then we need quite 
different methods in  order to find these. It seems that the  problem of the existence of the recursion 
operator and the integrability  of our systems is very tempting to study.

\end{document}